\documentclass[11pt]{article}
\usepackage[margin=1truein]{geometry}

\usepackage{setspace}
\usepackage{lscape}
\usepackage{pdflscape}

\usepackage{amsmath,amssymb,amsthm} 

\usepackage{wrapfig} 

\usepackage{mathrsfs} 
\usepackage{url} 
\usepackage{latexsym} 
\usepackage{bm} 

\usepackage{mathtools}

\usepackage{booktabs,multirow}
\usepackage{array}
\newcolumntype{P}[1]{>{\centering\arraybackslash}p{#1}}
\usepackage{diagbox}

\usepackage{algorithm}
\usepackage[noend]{algpseudocode}

\algnewcommand{\Break}{\textbf{break}}
\algnewcommand{\algorithmicgoto}{\textbf{go to} Line}
\algnewcommand{\Goto}[1]{\algorithmicgoto~\ref{#1}}
\algblockdefx{MRepeat}{EndRepeat}{\textbf{repeat}}{}
\algnotext{EndRepeat}

\usepackage[sort&compress,numbers]{natbib}

\usepackage{thmtools,thm-restate}

\usepackage{cleveref}
\expandafter\def\csname ver@etex.sty\endcsname{3000/12/31} 
\usepackage{autonum} 
\makeatletter
\autonum@generatePatchedReferenceCSL{Cref} 
\makeatother

\usepackage{tikz,tikz-3dplot}
\usetikzlibrary{positioning,shapes,shapes.callouts,shapes.multipart,arrows.meta,fit,calc}
\usepackage{pgfplots}
\pgfplotsset{compat=newest}
\usepackage{pxpgfmark} 
\usetikzlibrary{bayesnet} 

\usepackage{multicol}
\usepackage{wasysym} 
\usepackage{adjustbox}
\usepackage{xparse} 
\usepackage{pbox} 
\usepackage{authblk} 
\usepackage{xspace} 

\usepackage[subrefformat=parens]{subcaption}
\usepackage[inline]{enumitem} 

\usepackage{comment}

\DeclareMathOperator*{\argmax}{argmax}
\DeclareMathOperator*{\argmin}{argmin}

\DeclareMathOperator{\EE}{\mathbb{E}}

\DeclarePairedDelimiterX\inner[2]{\langle}{\rangle}{{#1},{#2}}
\DeclarePairedDelimiter\abs{|}{|}

\DeclarePairedDelimiter\set{\{}{\}}
\DeclarePairedDelimiter\prn{(}{)}
\DeclarePairedDelimiter\bra{[}{]}
\DeclarePairedDelimiterX\Set[2]{\{}{\}}{\mspace{2mu}{#1}\;\delimsize|\;{#2}\mspace{2mu}}
\DeclarePairedDelimiterX\Prn[2]{(}{)}{\mspace{2mu}{#1}\;\delimsize|\;{#2}\mspace{2mu}}
\DeclarePairedDelimiterX\Bra[2]{[}{]}{\mspace{2mu}{#1}\;\delimsize|\;{#2}\mspace{2mu}}

\newcommand{\R}{\mathbb R}

\newcommand{\1}{\mathbf 1}

\newcommand{\st}{\mathrm{s.t.}}

\renewcommand{\epsilon}{\varepsilon}

\NewDocumentCommand{\exsub}{s m O{} m}{%
  \IfBooleanT{#1}{\EE_{#2}\nolimits\bra*{#4}}%
  \IfBooleanF{#1}{\EE_{#2}\nolimits\bra[#3]{#4}}%
}

\newcommand{\dgamma}[2]{\gamma_{#1}(#2)}
\newcommand{\dgammaE}[2]{\gamma^{\mathrm{exp}}_{#1}(#2)}
\newcommand{\dgammaH}[2]{\gamma^{\mathrm{hyp}}_{#1}(#2)}
\newcommand{\dGamma}[2]{\Gamma_{#1}(#2)}

\newcommand{\mt}{\tilde{t}}

\newcommand{\diff}[1]{{#1}'}
\newcommand{\diffdiff}[1]{{#1}''}

\declaretheoremstyle[
]{thmsty}

\declaretheorem[
  name=Theorem,
  refname={Theorem,Theorems},
  style=thmsty,
]{theorem}
\declaretheorem[
  name=Corollary,
  refname={Corollary,Corollaries},
  style=thmsty,
]{corollary}
\declaretheorem[
  name=Proposition,
  refname={Proposition,Propositions},
  style=thmsty,
]{proposition}

\crefname{algorithm}{Algorithm}{Algorithms}
\crefname{line}{Line}{Lines}
\crefname{section}{Section}{Sections}
\crefname{appendix}{Appendix}{Appendices}
\crefname{table}{Table}{Tables}
\crefname{figure}{Figure}{Figures}
\crefname{equation}{}{}
\Crefname{equation}{Eq.}{Eqs.}

\setlist[itemize]{
  topsep=0.4\baselineskip,
  itemsep=0\baselineskip,
  leftmargin=1.5em,
}

\setlist[enumerate]{
  font=\upshape,
  label=(\alph*),
  ref=(\alph*),
  topsep=0.4\baselineskip,
  itemsep=0\baselineskip,
  leftmargin=2em,
}

\newlist{enuminasm}{enumerate}{1} 
\setlist[enuminasm]{
  font=\upshape,
  label=(\alph*),
  ref=\theassumption(\alph*),
  topsep=0.4\baselineskip,
  itemsep=0\baselineskip,
  leftmargin=2em,
}
\crefalias{enuminasmi}{assumption}

\newlist{enuminthm}{enumerate}{1}
\setlist[enuminthm]{
  font=\upshape,
  label=(\alph*),
  ref=\thetheorem(\alph*),
  topsep=0.4\baselineskip,
  itemsep=0\baselineskip,
  leftmargin=2em,
}
\crefalias{enuminthmi}{theorem}

\newlist{enuminlem}{enumerate}{1}
\setlist[enuminlem]{
  font=\upshape,
  label=(\alph*),
  ref=\thelemma(\alph*),
  topsep=0.4\baselineskip,
  itemsep=0\baselineskip,
  leftmargin=2em,
}
\crefalias{enuminlemi}{lemma}

\usepackage{datetime}
\date{\vspace{-2.5\baselineskip}}

\author[1]{Yasunori Akagi\footnote{Corresponding author. E-mail:yasunori.akagi@ntt.com}}
\author[1, 2]{Hideaki Kim}
\author[1]{Takeshi Kurashima}

\affil[1]{NTT Human Informatics Laboratories, Kanagawa, Japan}
\affil[2]{NTT Communication Science Laboratories, Kyoto, Japan}

\title{A Continuous-time Tractable Model for Present-biased Agents}

\begin{document}
\maketitle

\begin{abstract}
  Present bias, the tendency to overvalue immediate rewards while undervaluing future ones, is a well-known barrier to achieving long-term goals. As artificial intelligence and behavioral economics increasingly focus on this phenomenon, the need for robust mathematical models to predict behavior and guide effective interventions has become crucial. However, existing models are constrained by their reliance on the discreteness of time and limited discount functions. This study introduces a novel continuous-time mathematical model for agents influenced by present bias.
Using the variational principle, we model human behavior, where individuals repeatedly act according to a sequence of states that minimize their perceived cost.
Our model not only retains analytical tractability but also accommodates various discount functions. Using this model, we consider intervention optimization problems under exponential and hyperbolic discounting and theoretically derive optimal intervention strategies, offering new insights into managing present-biased behavior.
\end{abstract}

\section{Introduction} \label{sec: introduction}
Humans often fail to achieve their goals despite having long-term plans, even when no unforeseen events occur. Consider a person who plans to diet for a month. When making the plan, they expect to achieve their target weight by strictly managing their diet, believing they can resist temptations like a weekend invitation to a barbecue. However, as the weekend approaches, they succumb to temptation and go to the barbecue, failing their diet. Such behavior is referred to as \emph{time inconsistency} and is a significant topic in behavioral economics.

Time inconsistency is attributed to how humans discount future values, mainly due to \emph{present bias}. Present bias refers to the tendency to overvalue immediate costs and rewards while undervaluing those in the distant future. The mechanism by which present bias leads to time-inconsistent behavior can be explained using the example of a person planning a diet;
during the planning stage, the individual does not place a high value on the temptation of a weekend barbecue, which is still some time away, and believes they can resist it. However, as the weekend approaches, the value of the barbecue increases significantly due to present bias, making it much harder to resist the temptation.

Constructing behavior models incorporating present bias has become an important research topic in artificial intelligence and behavioral economics. It helps us predict human behavior and determine appropriate interventions for goal achievement. Especially, \citet{kleinberg2014time} introduced a behavior model under present bias using an agent that moves on a graph composed of vertices and edges. This model successfully replicates time-inconsistent behaviors studied extensively in behavioral economics, such as procrastination and task abandonment, while being applicable to various real-world tasks. 
Building on Kleinberg and Oren's model, \citet{akagi2023analytically} proposed a more feasible model restricting the tasks to \emph{progress-based tasks}. 
A progress-based task is one in which an agent accumulates a real-valued measure called \emph{progress} over a period and receives a reward if the progress reaches a predetermined goal. Common examples of such tasks in daily life include completing 20 hours of exercise within a week or finishing a graduation thesis within six months.
This model allows for the closed-form description of the agent's behavior and the efficient optimization of intervention, which were challenging in Kleinberg and Oren's model due to the high computational complexity \cite{tang2017computational,albers2019motivating}.

These models are crucial for examining the impact of present bias on human behavior, but they have two practical challenges when applied to real-world scenarios. 
One is that they are defined on a discretized time axis and demand the problematic predefinition of an appropriate step size of time (e.g., daily and weekly). 
If the time step is too large, the agent’s behavior can only be described in coarse granularity, while a too-small time step significantly increases the computational burden. 
The other is that the previous models only allow for a specific discount function called quasi-hyperbolic discounting \cite{laibson1997golden} with particular parameters to represent present bias. 
In behavioral economics, exponential discounting \cite{samuelson1937note} has traditionally been used as the classical discount function. 
However, to address time-inconsistent behavior that exponential discounting could not adequately capture, hyperbolic discounting \cite{ainslie1975specious} was introduced. Subsequently, quasi-hyperbolic discounting \cite{laibson1997golden} was proposed as an approximation to improve the tractability of hyperbolic discounting.
Previous models' inability to accommodate crucial discounting functions, such as exponential or hyperbolic discounting, poses significant constraints when analyzing real-world scenarios.

This paper introduces a new behavioral model for agents engaged in progress-based tasks under present bias in a continuous time setting.
Unlike previous models defined on a discretized time axis, the proposed model in a continuous time setting faces a non-trivial mathematical problem for describing and analyzing the agent behavior; the agent repeatedly selects the trajectory of their future state that minimizes the perceived cost at each point in time, makes small progress during \emph{an infinitesimal time interval}, and accumulates \emph{the infinite number of small progress} before reaching the goal progress. To deal with this problem properly, we formalize the agent’s behavior under present bias in terms of the variational principle, where the cost-minimizing trajectory of the agent solves a set of ordinary differential equations. The variational principle is a mathematical tool to treat the minimization/maximization problem of objective functionals, primarily utilized in physics \cite{berdichevsky2009variational} and machine learning \cite{kim2021fast,zappala2023neural}. Although the dynamics of an agent described above appear highly complex, we can obtain the agent’s trajectory in a very concise form under mild assumptions about discount functions. Notably, representative discount functions like exponential and hyperbolic discounting satisfy the assumptions, allowing the agent’s trajectory to be described in closed form or using easy-to-handle integrals.

Furthermore, we analyzed two optimization problems related to interventions using the derived closed-form trajectory equations. The first problem is the optimal goal-setting problem. This problem involves determining the progress goal that maximizes an individual's final progress when the period and reward are given. The problem formulation changes depending on whether permitting \emph{exploitative rewards}, which are decoy rewards that influence the agent's behavior. 
We analytically derived the optimal solutions for this problem when the discount function is either exponential or hyperbolic, considering both scenarios where exploitative rewards are allowed and not. Interestingly, we found that for both discount functions, the optimal solution remains unchanged regardless of whether we use exploitative rewards, indicating that exploitative rewards have no effect. 
This result contrasts with previous research in discrete-time settings with quasi-hyperbolic discounting, where exploitative rewards have been shown to have a significant impact \cite{akagi2023analytically}, suggesting a substantial difference between these discounting functions. 
Investigating which type of discount function the individual uses carefully is essential for better intervention; otherwise, there is a risk of applying a suboptimal intervention.

The second intervention optimization problem is the optimal reward scheduling problem. In this problem, given a total period and total reward, the goal is to divide the total period and decide rewards and progress targets for each sub-period to maximize the final progress achieved by the agent. We successfully derived the optimal solutions for this problem analytically in the cases of both exponential and hyperbolic discounting.
Our results show that when the number of divisions is fixed, it is optimal to divide the period into equal intervals, regardless of the parameters of the discount function. Moreover, the greater the number of divisions, the more significant the final progress can be achieved. Additionally, when the period is divided infinitely finely, the final progress converges to a limit independent of the shape of the discount function. This result suggests that dividing the period and rewards can effectively counteract the effects of present bias. These findings are also in clear contrast to those from discrete-time models with quasi-hyperbolic discounting; in such models, depending on the parameter values, it can sometimes be optimal not to divide the period into smaller intervals.

All proofs are deferred to the appendix.

\section{Related Work} \label{sec: related work}
Present bias and time-inconsistent behavior have long been significant research themes in behavioral economics \cite{frederick2002time,camerer2004behavioral,wilkinson2017introduction}. Present bias is modeled by time discount functions. Classical economic research has used exponential discounting \cite{samuelson1937note} as the time discount function, but to better explain human time-inconsistent behavior, which could not be adequately explained by exponential discounting alone, hyperbolic discounting \cite{ainslie1975specious} was introduced. Additionally, quasi-hyperbolic discounting \cite{laibson1997golden,phelps1968second} emerged as a discrete-time approximation of hyperbolic discounting. Other discounting functions have also been proposed, such as generalized hyperbolic functions \cite{loewenstein1992anomalies} and generalized Weibull functions \cite{takeuchi2011non}. In this study, we focus on exponential and hyperbolic discounting to explore the properties of the proposed model.

There has been extensive research on present bias, goal achievement, and incentives \cite{o2006incentives,o2008procrastination,koch2011self,hsiaw2013goal}.
Recent highlighted research on the mathematical modeling of goal achievement behavior under present bias is the graph-based model by \citet{kleinberg2014time}.
They model a task as a directed acyclic graph (DAG) and human behavior as agents moving through the DAG. 
The agent repeatedly calculates the cost for all paths to the goal, influenced by quasi-hyperbolic discounting, and proceeds along the path with the minimum cost.
Their model is flexible enough to handle various tasks, and they use it to examine the relationship between present bias and time inconsistency. 
Despite the novelty and superiority of their model, it also faced the challenge that solving the optimization problems for various interventions is often NP-Hard. \cite{albers2019motivating,albers2021value,tang2017computational}.

To address this issue, \citet{akagi2023analytically} proposed a new model based on the work of \citet{kleinberg2014time}. 
They demonstrated that the agent's behavior could be expressed in closed form by restricting their model to tasks where progress is accumulated in discrete time. Based on this formulation, they derived fast algorithms and closed-form optimal solutions for intervention optimization problems, such as optimal goal setting and reward scheduling, and analyzed their behavior. 
Their work dramatically inspires our research. Our contributions to their research are twofold.
Firstly, we demonstrated that even when extending their model to continuous time, the agent's trajectory can be expressed in a tractable form, allowing for a broader range of discount functions, thereby enhancing the model's applicability.
Secondly, similar to their study, we analyzed the conditions for task abandonment and the intervention optimization problem to clarify how the agent's behavior and optimal intervention strategies change due to differences between continuous and discrete time and differences in discount functions.

Our research is also closely related to reinforcement learning \cite{kaelbling1996reinforcement}. While our model focuses on explaining agents' behavior and determining optimal interventions under the assumption that agents maximize discounted rewards, reinforcement learning aims to teach agents how to act in unknown situations to achieve the same goal. Thus, our research and reinforcement learning explore the same issue from different angles. Additionally, research in reinforcement learning has extended time discount functions beyond exponential discounting \cite{schultheis2022reinforcement, fedus2019hyperbolic}, suggesting that future insights from these studies may intersect with our findings.

\section{Formulation of the Proposed Model} \label{sec: formulation}
\paragraph{Progress-based Task.}
Similar to the previous study \cite{akagi2023analytically}, this study deals with a task that aims to achieve a goal within a time limit by growing a real value called progress.
We assume that progress never decreases. 
Such types of tasks frequently occur in daily life.
For example, consider a student who aims to complete an assignment within 8 hours on a given day. If the student finishes the assignment, they will earn school credit for the course. In this case, the period corresponds to the 8 hours, progress corresponds to the percentage of the assignment completed, and the reward corresponds to the value of earning the course credit.
Another example could be a person participating in a health program that involves exercising for 30 hours within a month. In this scenario, the period is one month, and progress is measured by the hours spent exercising. The reward is the incentive provided upon completing the health program.

\begin{figure}[t]
  \centering
  \includegraphics[width=0.5\linewidth]{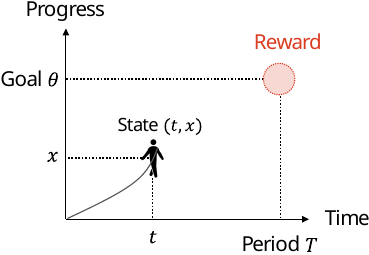}\hfill%
  \caption{An illustrative example of a progress-based task.} 
  \label{fig:task_decription}
\end{figure}

The agent is rewarded with $R$ if it achieves the goal progress $\theta$ within the period $T$, where $R, \theta, T \in \R_{\geq 0}$. 
We denote the agent's state by a tuple $(t, x)$, where $t \in [0, T]$ is the time and $x \in \R_{\geq 0}$ is the progress. 
Note that time $t$ is not restricted to discrete values but takes continuous values, a significant difference from the previous study \cite{akagi2023analytically}.
\cref{fig:task_decription} describes an illustrative example of a progress-based task.

\paragraph{Perceived Cost.}
Let the agent's state be $(t, x)$. 
We will denote by $\mathcal{Y}_{t, x}$ the set of absolutely continuous functions defined in $[t, T]$ which satisfies $y(t) = x$. 
A function $y(s) \in \mathcal{Y}_{t, x}$ represents the progress trajectory that an agent in state $(t,x)$ will take at future time $s \in [t, T]$.
For progress trajectory $y(s)$, we define \emph{perceived cost} $\mathcal{P}[y]$ by 
\begin{align} \label{eq:perceived cost}
 &\mathcal{P}[y] \coloneqq \int_{t}^T L_t(s, \diff{y}(s)) ds - \dgamma{t}{T} R \cdot \bm{1}[y(T) \geq \theta], \label{eq:perceived cost} \\
  &L_t(s, v) \coloneqq \dgamma{t}{s} c\prn*{v} \label{eq:L}, 
\end{align}
where $\diff{y}(s)$ denotes the derivative function of $y(s)$
Note that $\mathcal{P}[\cdot]$ is a \emph{functional} that takes a function as an argument and returns a real value.
The function $\gamma_t: \mathbb{R} \to (0, 1]$  is the \emph{discount function}, which represents how the agent discounts the future benefit or cost; an agent at time $t$ calculates the benefits and costs at a future time $s$ by multiplying the discount factor $\gamma_t(s)$.
We assume that $\gamma_t$ is continuously differentiable in $[t, T]$. 
The function $c:\R \to \R \cup \set*{+\infty}$ is the \emph{cost function}; to increase progress at a rate of $v$, a cost of $c(v)$ is incurred. 
Also, $\bm{1}[\cdot]$ is an indicator function; $\bm{1}[y_T \geq \theta] = 1$ if $y_T \geq \theta$, otherwise $\bm{1}[y_T \geq \theta] = 0$.

The intuitive interpretation of perceived cost \cref{eq:perceived cost} is the total cost of a future progress trajectory to an agent, considering the discount function effect. 
The first term of \cref{eq:perceived cost} is the total cost perceived by the agent over the period from $t$ to $T$, considering time discounting. It is calculated by integrating the agent's perceived cost $L_t(s, \diff{y}(s))$ at time $s$ from $t$ to $T$. 
The second is the reward term; the agent receives the time-discounted reward $\gamma_t(T)R$ if and only if the final progress $y(T)$ exceeds the goal progress $\theta$. 
This term has a negative sign because rewards have the opposite effect of costs. 

\begin{figure}[t]
  \centering
  \includegraphics[width=0.5\linewidth]{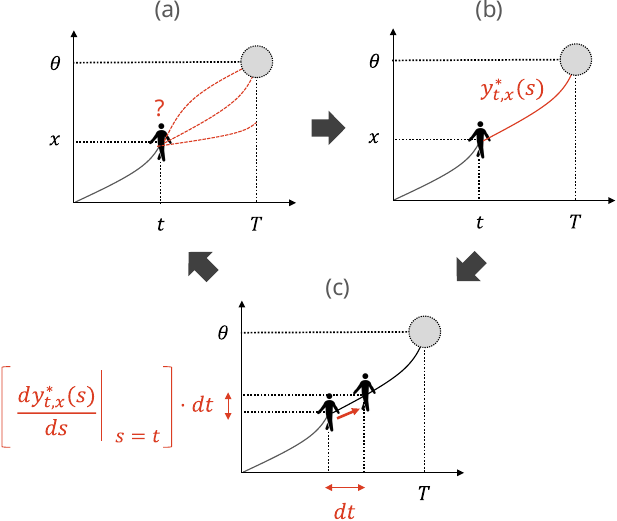}\hfill%
  \caption{
    An illustrative example of agents behavior model. 
    (a) The agent searches for a trajectory that minimizes the perceived cost among the candidate trajectories of future states $\mathcal{Y}_{t, x}$. The red dashed line is an example of a candidate trajectory in $\mathcal{Y}_{t, x}$.
    (b) Let $y^*_{t, x}(s)$ be the found trajectory that minimizes perceived cost.
    (c) The agent generates progress along $y^*_{t, x}(s)$. 
    Note that while the time interval $dt$ is inherently infinitesimal, it is depicted as larger for illustrative purposes.
  } 
  \label{fig:agent_behavior}
\end{figure}

\paragraph{Agent Behavior Model.}
The agent at state $(t, x)$ selects the trajectory $y^*_{t, x}$ with the minimum perceived cost among the candidate trajectories $\mathcal{Y}_{t, x}$. 
Then, at time $t$, the agent generates progress along $y^*_{t, x}$; it produces progress of 
$\prn*{\left. \frac{d y^*_{t, x}(s)}{d s} \right|_{s=t}} \cdot dt$
during the infinitesimal time $dt$. Starting from the state $(0, 0)$, the agent repeats this procedure at each time $t \in [0, T]$.
\cref{fig:agent_behavior} shows an illustrative example of this process. 

In this agent behavior model, the trajectory $x(t)$ actually taken by the agent is fully determined by \cref{eq:perceived cost,eq:L} and 
\begin{align}
  y^*_{t, x(t)} &\coloneqq \argmin_{y \in \mathcal{Y}_{t, x(t)}} \mathcal{P}[y], \label{eq:y^*} \\
  \frac{dx(t)}{dt} &= \left. \frac{d y^*_{t, x(t)}(s)}{ds} \right|_{s = t}, x(0) = 0. \label{eq:agent_ODE}
\end{align}

This agent's behavior model is a natural extension of the discrete-time behavior models from previous research \cite{kleinberg2014time,akagi2023analytically} to continuous time. In the previous models, the agent repeatedly chooses the path with the minimum perceived cost among all possible future paths at each time step and acts along that path.
In the model proposed in this paper, the path with the smallest perceived cost corresponds to $y^*_{t, x(t)}$ in \cref{eq:y^*} and acting along the path corresponds to \cref{eq:agent_ODE}. 

The formulation of the agent's behavior in \cref{eq:y^*} is based on the variational principle. The variational principle is a concept primarily used in physics and machine learning, stating that the state of a system is determined by a function that maximizes or minimizes a particular functional \cite{berdichevsky2009variational, kim2021fast, zappala2023neural}. In our formulation, the agent's future trajectory $y_{t, x(t)}^*$ is defined as the value that minimizes the functional $\mathcal{P}$, making it a clear example of the variational principle. By employing this formulation, we achieve a natural extension of existing agent behavior models to continuous time while also simplifying the analysis of agent behavior by utilizing the well-established variational methods techniques \cite{rockafellar2001convex}.

\begin{figure}[t]
  \centering
  \includegraphics[width=0.5\linewidth]{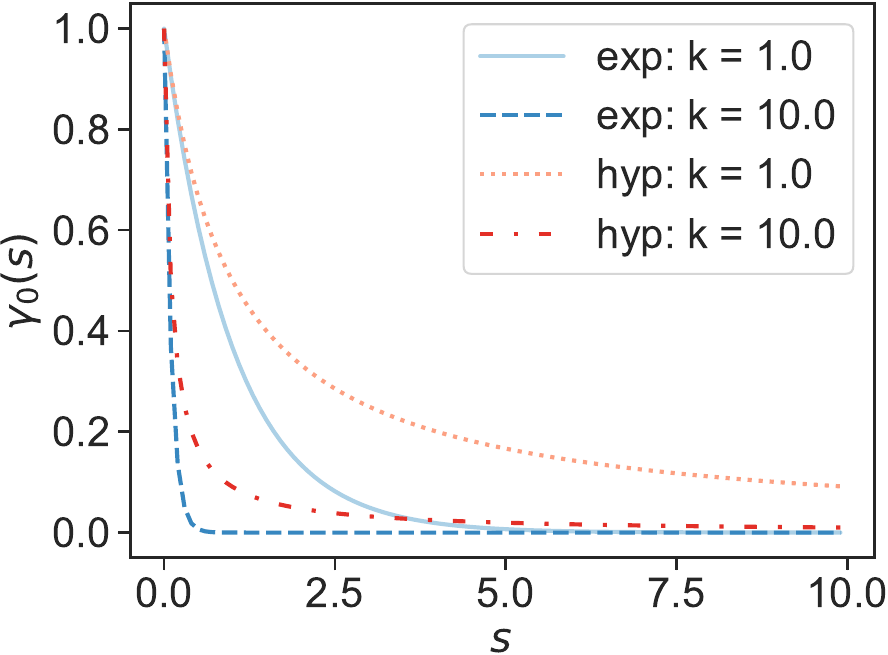}\hfill%
  \caption{The plots of exponential discount function $\dgammaE{0}{s}$ and hyperbolic discount function $\dgammaH{0}{s}$ for various parameters $k$. ``exp'' means exponential discount function and ``hyp'' means hyperbolic discount function. } 
  \label{fig:exp_vs_hyp}
\end{figure}

\paragraph{Discount Function.}
The discount function has a significant impact on the agent's behavior. 
In this paper, we focus on the most typical discount functions, the \emph{exponential discount function} \cite{samuelson1937note}
\begin{align} \label{eq:exponential_discounting}
    \dgammaE{t}{s} = \exp \prn*{- k (s - t)}
\end{align}
and the \emph{hyperbolic discount function} \cite{ainslie1975specious}
\begin{align} \label{eq:hyperbolic_discounting} 
    \dgammaH{t}{s} = \frac{1}{1 + k(s - t)},
\end{align}
where $k > 0$ is a parameter. 
Exponential discounting has been classically used as a human discounting model. 
Hyperbolic discounting was proposed to explain human behavior that exponential discounting cannot explain, such as time inconsistency.
For both discount functions, the function value decays rapidly when the parameter $k$ is large and slowly when it is small. Thus, we can interpret the magnitude of $k$ as the strength of the agent's present bias.
\cref{fig:exp_vs_hyp} shows the shapes of the exponential and hyperbolic discount functions. 

\paragraph{Cost Function.}
This paper assumes that the cost function can be written as 
\begin{align}
 \label{eq:cost function}
 c(\Delta)
 = 
 \begin{dcases*}
 \Delta^\alpha
 & if $\Delta \geq 0$,\\
 +\infty
 & otherwise, 
 \end{dcases*}
\end{align}
where $\alpha > 1$ is a parameter. 
Because this paper considers only tasks in which progress never decreases, we set $c(\Delta) = +\infty$ when $\Delta < 0$. 
We can control the shape of the cost function by adjusting the parameter $\alpha$.
This cost function is precisely the same as the one used in the previous study \cite{akagi2023analytically}.

\section{Properties of the Proposed Model}
\subsection{Formula of Agent's Trajectory} \label{subsec: trajectory formula}
The trajectory of the agent $x(t)$ is determined by \cref{eq:perceived cost,eq:L,eq:y^*,eq:agent_ODE}. 
This dynamics involves optimizing a functional (a variational problem) and differential equations, which are expected to result in very complex behavior. However, we have shown that under mild conditions, the agent's trajectory $x(t)$ can be expressed concisely.
For simplicity, we introduce the following notations; 
\begin{align}
  \Gamma_{t}(s) &\coloneqq \int \dgamma{t}{s}^{-\frac{1}{\alpha - 1}} ds, \label{eq:Gamma}\\
  \zeta(t) &\coloneqq \dgamma{t}{T} \prn*{\dGamma{t}{T} - \dGamma{t}{t}}^{\alpha - 1} \label{eq:zeta}. 
\end{align}
Note that \cref{eq:Gamma} is an indefinite integral; we can choose the constant of integration arbitrarily because it does not affect the later results. 

\begin{theorem} \label{thm:trajectory}
    Assume that $\zeta(t)$ is non-increasing in $[0, T]$. 
    The following holds:
    \begin{align} \label{eq:analytical_trajectory}
        x(t) = 
        \theta \prn*{1 - \exp \prn*{-\int_0^{\mt} \frac{ds}{\Gamma_{s}(T) - \Gamma_{s}(s)}}} \label{eq:x(t)}, 
    \end{align}
    where $\mt = \min (t, t^*)$ and $t^*$ is the smallest $t \in [0, T]$ such that
    \begin{align} \label{eq:abandonment_condition}
    \frac{\exp \prn*{- \alpha \int_0^{t} \frac{ds}{\Gamma_{s}(T) - \Gamma_{s}(s)}}}{\zeta(t)} &\geq \frac{R}{\theta^{\alpha}}
  \end{align}
  if there exists such $t$; otherwise, $t^* = T$. 
\end{theorem}

\cref{thm:trajectory} shows that we can calculate the agent's trajectory if the discount function $\gamma_t$ satisfies the property that $\Gamma_t(s)$ and $\int_0^{\mt} \frac{ds}{\Gamma_{s}(T) - \Gamma_{s}(s)}$ are analytically or numerically computable. 
The value $t^*$ is the \emph{abandonment time} when the agent gives up making further progress because any additional progress would not be worth the reward. 
The condition that $\zeta(t)$ is non-increasing corresponds to the scenario where once an agent abandons a task, they do not resume it (see \cref{subsec:proof of thm:trajectory}). If this condition does not hold, the agent would repeatedly abandon and resume the task, resulting in more complex behavior. While this case is also interesting, we leave a detailed analysis of it for future work, and this paper focuses on the case where $\zeta(t)$ is non-increasing. 

We explain the derivation of \cref{thm:trajectory} briefly. First, we use the variational method to solve for \cref{eq:y^*}. Typically, the variational method can only find local optima. However, in this case, due to the special structure of the cost function \cref{eq:cost function}, the functional $\mathcal{P}$ becomes a convex functional. This property allows us to obtain the global optimum  $y^*_{t, x(t)}$ by solving a differential equation called the Euler--Lagrange equation \cite{rockafellar2001convex}. 
We finally derive \cref{eq:analytical_trajectory} by solving the differential equation \cref{eq:agent_ODE} using the obtained $y^*_{t, x(t)}$. 
For details, please refer to \cref{subsec:proof of thm:trajectory}.

The discount function must satisfy several conditions to apply \cref{thm:trajectory}. 
Fortunately, the two leading discount functions, exponential and hyperbolic discounting, satisfy these conditions. 

\begin{corollary} \label{thm:exp_trajectory}
    When $\gamma_t = \gamma^{\mathrm{exp}}_t$ defined as \cref{eq:exponential_discounting}, $\zeta(t)$ is non-increasing and 
    \begin{align}
        x(t) &= \theta \prn*{\frac{\exp \prn*{\frac{k}{\alpha-1} \mt} - 1}{\exp \prn*{\frac{k}{\alpha-1}T} - 1}} \label{eq:x(t)_exp}, \\
        t^* &=
        \begin{cases}
        0 & \prn*{\frac{k}{(\alpha - 1) \prn*{1 - \exp \prn*{-\frac{kT}{\alpha-1}}}}}^{\alpha-1} \geq \frac{R}{\theta^{\alpha}}, \\
        T & \mathrm{otherwise}. 
        \end{cases} \label{eq:exp_abandonment_condition}
    \end{align}
\end{corollary}
\begin{corollary} \label{thm:hyp_trajectory}
    When $\gamma_t = \gamma^{\mathrm{hyp}}_t$ defined as \cref{eq:hyperbolic_discounting}, $\zeta(t)$ is non-increasing, 
    \begin{align}
            x(t) &= \theta \prn*{1 - \exp \prn*{- \frac{k \alpha}{\alpha-1} \eta(\mt)}}, \label{eq:x(t)_hyp} \\
            \eta(t) &\coloneqq \int_{0}^{t} \frac{ds}{\prn*{1 + k(T - s)}^{\frac{\alpha}{\alpha-1}} - 1}, 
    \end{align}
    and $t^*$ is the smallest $t \in [0, T]$ such that 
    \begin{align}
        \prn*{\frac{k \alpha}{(\alpha-1) \prn*{\prn*{1 + k(T-t)}^{\frac{\alpha}{\alpha-1}}-1}}}^{\alpha-1} \prn*{1 + k(T-t)} \exp \prn*{-\frac{k \alpha^2}{\alpha-1} \eta(t)} \geq \frac{R}{\theta^{\alpha}} \label{eq:hyp_abandonment_condition}
    \end{align}
    if there exists such $t$; otherwise, $t^* = T$. 
\end{corollary}
\begin{corollary} \label{thm:hyp_2_trajectory}
    When $\gamma_t = \gamma^{\mathrm{hyp}}_t$ and $\alpha=2$, 
    \begin{align}
        x(t) = \frac{2 \theta \mt}{\prn*{k (T - \mt) + 2}T} \label{eq:x(t)_hyp_2}
    \end{align}
    and $t^*$ is the smallest $t \in [0, T]$ such that 
    \begin{align}
        \frac{2(kT+2)^2}{T^2} \frac{(T-t) \prn*{k(T-t) + 1}}{\prn*{k(T-t)+2}^3} \geq \frac{R}{\theta^{2}} \label{eq:hyp_2_abandonment_condition}
    \end{align}
    if there exists such $t$; otherwise, $t^* = T$. 
\end{corollary}
\cref{thm:exp_trajectory,thm:hyp_trajectory,thm:hyp_2_trajectory} indicate that we can express the agent's trajectory in closed form for the cases of exponential discounting and hyperbolic discounting with $\alpha = 2$. 
For hyperbolic discounting with $\alpha \neq 2$,  we cannot derive a closed-form expression of the agent's trajectory without integration; however, we can calculate it by a simple numerical integration of a single variable.

One of the proposed model's advantages is that the agent's trajectory can be obtained in a tractable form for representative discount functions. In the subsequent sections, we will use these tractable formulas to investigate the properties of agent behavior with present bias in continuous time, including task abandonment and optimal interventions.

\subsection{Abandonment Time and Time-inconsistency} \label{subsec:abandonment_time}
Here, we will examine the task abandonment time $t^*$ and the conditions under which the agent abandons the task, exploring the circumstances under which time-inconsistent behavior occurs.

First, consider the case of exponential discounting. According to \cref{thm:exp_trajectory}, the abandonment time $t^*$ takes either the value of $0$ or $T$. Therefore, the agent either completes the task or does not engage with the task at all from the beginning without making any progress. In other words, an agent with exponential discounting never abandons a task once started. In this sense, an agent with exponential discounting is time-consistent.

Next, consider the case of hyperbolic discounting ($\alpha = 2$ for simplicity).
\begin{proposition} \label{thm:hyp_2_abandonment}
    If $T \leq \frac{1 + \sqrt{3}}{k}$, the left-hand side of \cref{eq:hyp_2_abandonment_condition} takes maximum value $\frac{2(kT+1)}{kT+2}$ at $t = 0$. 
    Otherwise, the left-hand side of \cref{eq:hyp_2_abandonment_condition} takes maximum value $\frac{(kT+2)^2}{3 \sqrt{3} kT^2}$ at $t = T - \frac{1 + \sqrt{3}}{k}$. 
\end{proposition}
\cref{thm:hyp_2_abandonment} indicates that if $T \leq \frac{1 + \sqrt{3}}{k}$, $t^*$ takes either the value of $0$ or $T$ and so the agent's behavior is time-consistent as the exponential discounting case. On the other hand, if $T > \frac{1 + \sqrt{3}}{k}$, $t^*$ may take a value other than $0$ or $T$, because the left-hand side of \cref{eq:hyp_2_abandonment_condition} takes maximum value at $t =  T- \frac{1 + \sqrt{3}}{k}$. This implies that an agent may initially perceive the task as beneficial and start working on it, but later, reconsidering the costs and benefits, decide that continuing the task is disadvantageous and abandon it midway. This demonstrates the possibility of time-inconsistent behavior, where the agent's preferences change over time. 

It is well-known in behavioral economics that the behavior of individuals with exponential discounting is time-consistent, while that of individuals with hyperbolic discounting is time-inconsistent \cite{frederick2002time}. The results in this section demonstrate that our model successfully reproduces these insights in behavioral economics.

\section{Optimal Goal-Setting Problem} \label{sec:optimal_goal_setting}
In this section, we consider the problem of finding the goal $\theta^*$ which maximizes the final progress $x(T)$:
\begin{align}
  \theta^* \coloneqq \argmax_{\theta \geq 0}\ x(T),
  \label{eq:problem_goal_setting}
\end{align}
given the period $T$, reward $R$, and parameters $k$ and $\alpha$. 
This problem involves determining how an intervener should set goals to maximize an agent's progress over a fixed period with a given reward.

This problem is crucial in real-world scenarios. For instance, consider a company president (intervener) who offers employees (agents) a fixed bonus for achieving a sales target within six months. How should the president set this target to maximize final sales? If the target is set too high, employees may lose motivation due to the difficulty of achieving the target, resulting in lower final sales. On the other hand, if the target is too low, employees might reach it easily, earn the bonus, and feel satisfied, resulting in reduced effort and lower sales. Considering this trade-off, setting appropriate goals, as formulated in \cref{eq:problem_goal_setting}, is critical.

The optimal solution to the goal-setting problem depends on whether exploitative rewards are allowed. Exploitative rewards are decoys that take advantage of an agent's present bias. An agent with present bias may make progress even if they ultimately abandon the goal. As a result, setting high, unattainable goals can sometimes lead to more significant progress than setting lower, achievable goals, even though the agent fails to reach the final goal. 
This approach also benefits the intervener, because he does not need to give the reward because the agent does not achieve the goal. However, exploitative rewards raise ethical concerns, such as reducing motivation and hindering long-term effort, requiring careful consideration. Existing research has discussed the significance and propriety of exploitative rewards \cite{kleinberg2014time,tang2017computational,albers2019motivating}. A particularly important result is that using exploitative rewards for agents with quasi-hyperbolic discounting can increase overall progress significantly \cite{akagi2023analytically}.

This section examines both scenarios: allowing and disallowing exploitative rewards. The optimization problem when exploitative rewards are permitted is precisely \cref{eq:problem_goal_setting}, while the problem when they are not allowed adds the constraint $x(T) = \theta$ to \cref{eq:problem_goal_setting}. 
The following theorems give the optimal solutions to the optimal goal-setting problems. 

\begin{theorem} \label{thm:exp_optimal_goal}
When $\gamma_t = \gamma_t^{\mathrm{exp}}$, regardless of whether exploitative rewards are permitted, 
\begin{align}
    \theta^* = \bra*{\frac{\alpha-1}{k} \prn*{1 - \exp \prn*{-\frac{kT}{\alpha-1}}}}^{\frac{\alpha-1}{\alpha}} R^{\frac{1}{\alpha}}
\end{align}
and $\max_{\theta \geq 0}\ x(T) = \theta^*$. 
\end{theorem}

\begin{theorem} \label{thm:hyp_optimal_goal}
When $\gamma_t = \gamma_t^{\mathrm{hyp}}$ and $\alpha=2$, regardless of whether exploitative rewards are permitted, 
    \begin{align}
        \theta^* = 
        \begin{cases}
        \frac{T \prn*{3 \sqrt{3} k}^{\frac{1}{2}}}{kT+2} R^{\frac{1}{2}} & T \geq \frac{1 + \sqrt{3}}{k}, \\
        \prn*{\frac{T (kT+2)}{2(kT+1)}}^{\frac{1}{2}} R^{\frac{1}{2}} & T < \frac{1 + \sqrt{3}}{k}
        \end{cases}
    \end{align}
    and $\max_{\theta \geq 0}\ x(T) = \theta^*$. 
\end{theorem}

From \cref{thm:exp_optimal_goal} and \cref{thm:hyp_optimal_goal}, we observe followings. 
\paragraph{Effect of $k$.} When $T$ and $R$ are fixed, $\theta^*$ is monotonically decreasing with respect to $k$ for both discounting functions, and $\lim_{k \to \infty} \theta^* = 0$. 
This property implies that the more rapidly the discount function decays, the smaller the achievable goal progress.
\paragraph{Effect of $T$.} When $k$ and $R$ are fixed, $\theta^*$ is monotonically increasing with respect to $T$ for both discounting functions, and $\lim_{T \to \infty} \theta^* = \prn*{\frac{\alpha-1}{k}}^{\frac{\alpha-1}{k}} R^{\frac{1}{\alpha}}$ with exponential discounting and $\lim_{T \to \infty} \theta^* = \prn*{3 \sqrt{3}/k}^{\frac{1}{2}} R^{\frac{1}{2}}$ with hyperbolic discounting. 
This property implies that while longer time limits lead to higher achievable goals, there is an upper bound on the attainable goals. 
\paragraph{Effect of Exploitative Reward.} For both discount functions, whether we permit exploitative rewards or not does not change $\theta^*$. This result means that leveraging present bias cannot exploit the agents with these discount functions. 
This result contrasts with the significant impact of exploitative rewards under quasi-hyperbolic discounting, which approximates hyperbolic discounting \cite{akagi2023analytically}. It highlights a substantial difference in the optimal interventions for hyperbolic versus quasi-hyperbolic discounting. Therefore, when designing interventions, it is crucial to determine the type of discount function the individual uses accurately; otherwise, the derived intervention may be suboptimal.

\section{Optimal Reward Scheduling Problem}
This section considers an advanced intervention optimization problem, optimal reward scheduling problem. In this problem, the intervener can divide the rewards and distribute them to the agent multiple times.

The inputs to the problem are the total period $T$, the total reward $R$, and the agent parameters $k$ and $\alpha$.
The intervener choose some positive integer $N$ and divide the total period $T$ into $N$ periods $T_1, \ldots, T_N \in \R_{\geq 0}$ and the total reward $R$ into $N$ rewards $R_1, \ldots, R_N \in \R_{\geq 0}$.
Note that $\sum_{i=1}^N T_i = T$ and $\sum_{i=1}^N R_i = R$ must hold. 
The intervener also decide the goal progress $\theta_i \in \R_{\geq 0}$ for $i \in \set*{1, \ldots, N}$. 
We refer to the tuple $\prn*{N, (T_i)_{i=1}^N, (R_i)_{i=1}^N, (\theta_i)_{i=1}^N}$ as reward schedule. 

\begin{figure}[t] 
  \centering 
  \includegraphics[width=0.5\linewidth]{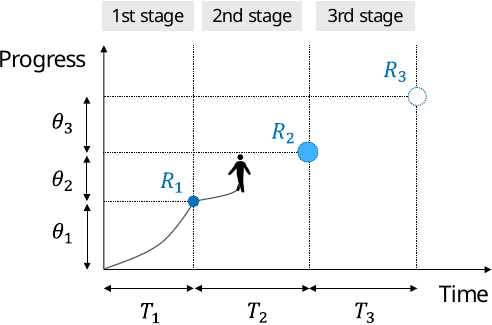}\hfill%
  \caption{An example of reward schedule when $N=3$. } 
  \label{fig:reward schedule}
\end{figure}

Once the reward schedule is determined, the agent accumulates progress in $N$ stages.
In the $i$-th stage, agents accumulate progress under the condition that they will receive a reward $R_i$ if they achieve the goal progress $\theta_i$ in the period $T_i$. 
Note that the agent makes independent progress at each stage. In other words, the agent makes progress at the $i$-th stage without considering the rewards or goals at stages other than the $i$-th. Let $P_i$ be the increment of progress in the $i$-th stage. We aim to maximize the overall increment of progress $\sum_{i=1}^{N} P_i$ by appropriately deciding the reward schedule.
\cref{fig:reward schedule} illustrates an example of a reward schedule in this setting.

\begin{theorem} \label{thm:orsp_fixed_N_exp}
    When $\gamma_t = \gamma_t^{\mathrm{exp}}$, for fixed $N$, the optimal schedule is 
    \begin{align}
        T_i &= \frac{T}{N},
        R_i = \frac{R}{N}, \\
        \theta_i &= \bra*{\frac{\alpha-1}{k} \prn*{1 - \exp \prn*{-\frac{kT}{(\alpha-1)N}}}}^{\frac{\alpha-1}{\alpha}} \prn*{\frac{R}{N}}^{\frac{1}{\alpha}}
    \end{align}
    for all $i \in \set*{1, \ldots, N}$, and the objective function value is 
    \begin{align}
        &f_{\alpha}^{\mathrm{exp}}(N) \coloneqq \\
        &\prn*{\frac{N(\alpha-1)}{k} \prn*{1 - \exp \prn*{-\frac{kT}{(\alpha-1)N}}}}^{\frac{\alpha-1}{\alpha}} R^{\frac{1}{\alpha}}. \label{eq:optimal_opt_exp} 
    \end{align}
\end{theorem}

\begin{theorem} \label{thm:orsp_fixed_N_hyp}
    When $\gamma_t = \gamma_t^{\mathrm{hyp}}$ and $\alpha = 2$, for fixed $N$, the optimal schedule is \begin{align}
        T_i &= \frac{T}{N}, \ \ 
        R_i = \frac{R}{N}, \\
        \theta_i &=
        \begin{cases}
        \frac{T \prn*{3 \sqrt{3} k}^{\frac{1}{2}}}{kT+2N} \prn*{\frac{R}{N}}^{\frac{1}{2}} & \frac{T}{N} \geq \frac{1 + \sqrt{3}}{k}, \\
        \prn*{\frac{T (kT+2N)}{2N(kT+N)}}^{\frac{1}{2}} \prn*{\frac{R}{N}}^{\frac{1}{2}} & \frac{T}{N} < \frac{1 + \sqrt{3}}{k},
        \end{cases}
    \end{align}
    for all $i \in \set*{1, \ldots, N}$, and the objective function value is 
    \begin{align} \label{eq:optimal_opt_hyp} 
    f_2^{\mathrm{hyp}}(N) \coloneqq
        \begin{cases}
        \frac{T \prn*{3 \sqrt{3} Nk}^{\frac{1}{2}}}{kT+2N} R^{\frac{1}{2}} & \frac{T}{N} \geq \frac{1 + \sqrt{3}}{k}, \\
        \prn*{\frac{T (kT+2N)}{2(kT+N)}}^{\frac{1}{2}} R^{\frac{1}{2}} & \frac{T}{N} < \frac{1 + \sqrt{3}}{k}. 
        \end{cases}
    \end{align}
\end{theorem}

\begin{theorem} \label{thm:orsp_N_exp}
    $f_{\alpha}^{\mathrm{exp}}(N)$ defined by \cref{eq:optimal_opt_exp} is monotonically increasing with respect to $N$. Furthermore, $f_{\alpha}^{\mathrm{exp}}(N) < T^{\frac{\alpha-1}{\alpha}}R^{\frac{1}{\alpha}}$ and $\lim_{N \to \infty} f_{\alpha}^{\mathrm{exp}}(N) = T^{\frac{\alpha-1}{\alpha}}R^{\frac{\1}{\alpha}}$ hold. 
\end{theorem}

\begin{theorem} \label{thm:orsp_N_hyp}
    $f_{2}^{\mathrm{hyp}}(N)$ defined by \cref{eq:optimal_opt_hyp} is monotonically increasing with respect to $N$. Furthermore, $f_{2}^{\mathrm{hyp}}(N) < T^{\frac{1}{2}}R^{\frac{1}{2}}$ and $\lim_{N \to \infty} f_{2}^{\mathrm{exp}}(N) = T^{\frac{1}{2}}R^{\frac{1}{2}}$ hold. 
\end{theorem}

\cref{thm:orsp_fixed_N_exp,thm:orsp_fixed_N_hyp} imply that when we fix $N$, we can write the optimal reward schedule in closed form. Furthermore, for both exponential and hyperbolic discounting, it is optimal to evenly distribute the total rewards and provide them at regular intervals. 
\cref{thm:orsp_N_exp,thm:orsp_N_hyp} indicate that the more times we divide the rewards or periods, the more significant the total progress the agent ultimately achieves. However, there is an upper bound to the total progress, and as $N$ increases, the total progress converges to this upper bound.

An important observation is that the limit is independent of the form of the discount function. For example, the limit of $f_{\alpha}^{\mathrm{exp}}(N)$, $T^{\frac{\alpha-1}{\alpha}}R^{\frac{\1}{\alpha}}$ and the limit of $f_{2}^{\mathrm{exp}}(N)$, $T^{\frac{1}{2}}R^{\frac{1}{2}}$ do not depend on the parameter $k$ that changes the shape of the discount function. Furthermore, both $f_{2}^{\mathrm{exp}}(N)$ and $f_{2}^{\mathrm{hyp}}(N)$ converge to the same value $T^{\frac{1}{2}} R^{\frac{1}{2}}$. 
These results imply that providing intermediate rewards at sufficiently fine intervals can counteract the effects of present bias and achieve progress independent of the individual's discount function.
In practical situations, it is impossible to divide rewards infinitely finely, and there are often constraints on the number of divisions. However, even in such cases, splitting the rewards as finely as possible can mitigate the effects of present bias and maximize the attainable progress.

This result contrasts with the previous findings under quasi-hyperbolic discounting; \citet{akagi2023analytically} demonstrated that under discrete time quasi-hyperbolic discounting, varying the reward intervals based on the discount-function parameter is optimal. For specific parameters, providing the rewards in a lump sum is optimal, whereas for other parameters, dividing the rewards as finely as possible is optimal. In contrast, in the continuous-time model proposed in this paper, the optimal strategy is to divide the rewards as finely as possible, regardless of the type (exponential or hyperbolic) and parameters of the discount function. The reasons for this difference may include the distinctions between discrete and continuous time and differences in the shapes of the discount functions (exponential and hyperbolic discount functions change smoothly, whereas quasi-hyperbolic discounting changes discontinuously). We will leave more detailed explanations in future work.

\section{Conclusion}
This paper introduces a new mathematical model for human goal achievement behavior under present bias in continuous time. The model is analytically tractable and accommodates essential discount functions, such as exponential and hyperbolic discounting. With this model, we derived analytical conditions for agents abandoning tasks and identified optimal intervention strategies, leading to valuable insights. Notably, our findings differ from those obtained using quasi-hyperbolic discounting in discrete time, emphasizing the importance of selecting the appropriate model based on the specific context.

\bibliographystyle{abbrvnat}
\bibliography{main}

\clearpage
\appendix
\section{Existing Results in Variational Methods}
Here, we provide an overview of the existing results in variational methods related to our paper.
We use the contents presented here in \cref{subsec:proof of thm:trajectory}. 

Let us denote by $\mathcal{A}^p[a, b]$ the set of functions that are absolutely continuous on the interval $[a, b]$ and whose derivatives are $p$-integrable, where $p \in [1, \infty]$. 
We consider the following variational problem; 
\begin{align}
    \label{eq:variational problem strict}
    \begin{aligned}
    \underset{y \in \mathcal{A}^p}{\mathrm{min.}}& \quad \int_{a}^{b} f(s, y, \diff{y}(x)) ds \\
    \mathrm{s.t.}& \quad y(a) = A, \ y(b) = B, 
    \end{aligned}
\end{align}
where $f(x_0, x_1, x_2)$ is continuously differentiable. 
Write $f_i(x_0, x_1, x_2) = \frac{\partial}{\partial x_i} f(x_0, x_1, x_2)\ (i=0, 1, 2)$. 
Then, the following theorem holds. 

\begin{theorem}[\cite{rockafellar2001convex}, Theorem 3] \label{thm:thm:variational_optimality_strict}
If the function $f(x_0, x_1, x_2)$ is convex with respect to $(x_1, x_2)$ and there exist functions $y \in \mathcal{A}^p[a, b]$ and $z \in \mathcal{A}^1[a, b]$ that satisfy
\begin{align}
    &z(s) = f_2(s, y, \diff{y}),\ z'(x) =f_1(s, y, \diff{y}), \label{eq:Euler-Lagrange_strict} \\
    &y(a) = A,\ y(b) = B, 
\end{align}
then $y$ is an optimal solution of \cref{eq:variational problem strict}. 
\end{theorem}
The equations \cref{eq:Euler-Lagrange_strict} are called Euler--Lagrange equation. 

\section{Proofs}

\subsection{Proof of \cref{thm:trajectory}} \label{subsec:proof of thm:trajectory}
\begin{proof}
First, we solve \cref{eq:y^*}. 
A function $y$ that minimizes $\mathcal{P}[y]$ satisfies $y(T) = \theta$ or $y(T)=x(t)$. 
This is because $y(T) = \theta$ holds when $y(T) \geq \theta$ and $y(T) = x(t)$ holds otherwise from the definitions of functional $\mathcal{P}[y]$ and cost function $c(\Delta)$. 

(i) When $y(T) = x(t)$, it is clear that the optimal choice is $y(s) = x(t)$ for $s \in [t, T]$, and in this case, $\mathcal{P}[y]=0$. 

(ii) When $y(T) = \theta$, because 
\begin{align}
    \mathcal{P}[y] = \int_{t}^T L_t(s, \diff{y}(s)) ds - \dgamma{t}{T} R 
\end{align}
holds, it is sufficient to solve the problem 
\begin{align}
    \min_{y \in \mathcal{A}^1[t, T]} \int_{t}^T L_t(s, \diff{y}(s)) ds \quad \mathrm{s.t.}\  y(t) = x(t),\  y(T) = \theta. \label{eq:variational problem}
\end{align}
Note that $\mathcal{A}^1[t, T]$ consists of all absolutely continuous function in $[t, T]$.  

Here, we consider a new variational problem 
\begin{align}
    \min_{y \in \mathcal{A}^1[t, T]} \int_{t}^T \tilde{L}_t(s, \diff{y}(s)) ds \quad \mathrm{s.t.}\  y(t) = x(t),\  y(T) = \theta, \label{eq:variational problem tilde}
\end{align}
where 
\begin{align}
    \tilde{c}(\Delta) &\coloneqq \abs{\Delta}^{\alpha}, \\
    \tilde{L}_t(s, v) &\coloneqq \gamma_t(s) \tilde{c}(v). 
\end{align}
In this problem, if an optimal solution exists for problem \cref{eq:variational problem tilde}, then an optimal solution also exists for problem \cref{eq:variational problem}, and the optimal solutions are identical; this is because the optimal solution to problem \cref{eq:variational problem tilde} must be non-increasing due to the form of $\tilde{c}(\Delta)$. 
Therefore, instead of solving problem \cref{eq:variational problem}, we consider solving problem \cref{eq:variational problem tilde}. 
The function $\tilde{L}_t(s, \diff{y}(s)) = \gamma_t(s) \abs{\diff{y}(s)}^{\alpha}$ is continuously differentiable and convex with respect to $(y(s), \diff{y}(s))$. Also, 
\begin{align}
    y(s) &= x(t) + (\theta - x(t)) \cdot \frac{\Gamma_t(s) - \Gamma_t(t)}{\Gamma_t(T) - \Gamma_t(t)} \label{eq:optimal_y}, \\
    z(s) &= \alpha \prn*{\frac{\theta - x(t)}{\Gamma_t(T) - \Gamma_t(t)}}^{\alpha-1}
\end{align}
satisfy the Euler--Lagrange equation \cref{eq:Euler-Lagrange_strict} and $y, z \in \mathcal{A}^1[a, b]$. 
From \cref{thm:thm:variational_optimality_strict}, \cref{eq:variational problem tilde} has an optimal solution, and the solution is given by \cref{eq:optimal_y}. 
Thus, \cref{eq:optimal_y} is also an optimal solution of \cref{eq:variational problem}
, and in this case a simple calculation gives 
\begin{align}
    \mathcal{P}[y] &=  \frac{(\theta - x(t))^{\alpha} - \zeta(t) R }{(\Gamma_t(T) - \Gamma_t(t))^{\alpha-1}}. 
\end{align} 

From (i) and (ii), we have
\begin{align} \label{eq:y^*_solved}
    y^*_{t, x(t)}(s) =
    \begin{cases}
         x(t) & (\theta - x(t))^{\alpha} - \zeta(t) R \geq 0, \\
        x(t) +  \frac{\prn*{\theta - x(t)} \prn*{\Gamma_t(s) - \Gamma_t(t)}}{\Gamma_t(T) - \Gamma_t(t)} & \mathrm{otherwise}. 
    \end{cases}
\end{align}

Next, we solve \cref{eq:agent_ODE}. By differentiating \cref{eq:y^*_solved}, we get 
\begin{align}
    \frac{d y^*_{t, x(t)}(s)}{ds} = 
    \begin{cases}
         0 & (\theta - x(t))^{\alpha} - \zeta(t) R \geq 0, \\
        (\theta - x(t)) \cdot \frac{\gamma_t(s)^{-\frac{1}{\alpha-1}}}{\Gamma_t(T) - \Gamma_t(t)} & \mathrm{otherwise}.
    \end{cases}
\end{align}
If the condition $(\theta - x(\hat{t}))^{\alpha} - \zeta(\hat{t}) R \geq 0$ is satisfied for a certain $\hat{t}$, then it follows from the form of \cref{eq:agent_ODE} and non-increasingness of $\zeta(t)$, that $x(t)$ remains constant for $t \in \bra*{\hat{t}, T}$. 
Let $\mt$ denote the smallest such $\hat{t}$. 
Then, 
\begin{align}
    \frac{d y^*_{t, x(t)}(s)}{ds} = (\theta - x(t)) \cdot \frac{\gamma_t(s)^{-\frac{1}{\alpha-1}}}{\Gamma_t(T) - \Gamma_t(t)} \label{eq:agent_ODE_simple}
\end{align}
holds for $t \in \bra*{0, \mt}$. 
In this case, solving \cref{eq:agent_ODE_simple} under the condition $x(t) = 0$ yields
\begin{align}
    x(t) = 
    \theta \prn*{1 - \exp \prn*{-\int_0^{t} \frac{ds}{\Gamma_{s}(T) - \Gamma_{s}(s)}}}. 
\end{align}
Because
\begin{align}
    (\theta - x(t))^{\alpha} - \zeta(t) R =
    \theta^{\alpha} \exp \prn*{-\alpha \int_0^{t} \frac{ds}{\Gamma_{s}(T) - \Gamma_{s}(s)}} - \zeta(t) R
\end{align}
holds, $\mt$ is the minimum $t$ that satisfies \cref{eq:abandonment_condition}. 
\end{proof}
\subsection{Proof of \cref{thm:exp_trajectory}}
\begin{proof}
When $\gamma_t = \gamma^{\mathrm{exp}}_t$, we have
\begin{align}
    \Gamma_t(s) &= \frac{\alpha-1}{k} \exp \prn*{\frac{k(s-t)}{\alpha-1}}, \label{eq:Gamma_exp}\\
    \zeta(t) &= \prn*{\frac{\alpha-1}{k}}^{\alpha-1} \prn*{1 - \exp \prn*{-\frac{k(T-t)}{\alpha-1}}}^{\alpha-1}. \label{eq:zeta_exp}
\end{align}
Obviously, $\zeta(t)$ is non-increasing in [0, 1]. 
By substituting \cref{eq:Gamma_exp} into \cref{eq:analytical_trajectory}, we get \cref{eq:x(t)_exp}. 
To show \cref{eq:exp_abandonment_condition}, we substitute \cref{eq:Gamma_exp,eq:zeta_exp} into \cref{eq:abandonment_condition} and get 
\begin{align}
    \prn*{\frac{k}{\alpha-1}}^{\alpha-1} \frac{\exp(kT)}{\prn*{\exp \prn*{\frac{k}{\alpha-1}T}-1}^{\alpha}} \prn*{\exp \prn*{\frac{k}{\alpha-1}T} - \exp \prn*{\frac{k}{\alpha-1}t} }  \geq \frac{R}{\theta^{\alpha}}. \label{eq:exp_abandonment_condition_trans} 
\end{align}
Because the left-hand side of \cref{eq:exp_abandonment_condition_trans} is monotonically decreasing with respect to $t$, $t^* = 0$ if \cref{eq:exp_abandonment_condition_trans} holds with $t=0$ and $t^* = T$ otherwise. 
\end{proof}

\subsection{Proof of \cref{thm:hyp_trajectory}}
\begin{proof}
When $\gamma_t = \gamma^{\mathrm{hyp}}_t$, we have
\begin{align}
    \Gamma_t(s) &= \frac{\alpha - 1}{k \alpha} \prn*{1 + k(s-t)}^{\frac{\alpha}{\alpha-1}}, \label{eq:Gamma_hyp}\\
    \zeta(t) &= \prn*{\frac{\alpha-1}{k\alpha}}^{\alpha-1} \frac{\set*{\prn*{1+k(T-t)}^{\frac{\alpha}{\alpha-1}}-1}^{\alpha-1}}{1 + k(T - t)}. \label{eq:zeta_hyp}
\end{align}
By differentiating \cref{eq:zeta_hyp} with $t$, we can show that $\zeta(t)$ is non-increasing for $t \in [0, T]$. 
By substituting \cref{eq:Gamma_hyp,eq:zeta_hyp} into \cref{eq:analytical_trajectory,eq:abandonment_condition}, we get \cref{eq:x(t)_hyp,eq:hyp_abandonment_condition}. 
\end{proof}

\subsection{Proof of \cref{thm:hyp_2_trajectory}}
\begin{proof}
When $\alpha = 2$, 
\begin{align}
\eta(t) &= \int_0^{\tau} \frac{1}{\prn*{1 + k(T - t)}^2 - 1} dt \\
&= \frac{1}{2 k} \prn*{\log \frac{k T}{ k T + 2} - \log \frac{k (T - \tau)}{ k (T - \tau) + 2}} \label{eq:eta_2} 
\end{align}
holds. By substituting \cref{eq:eta_2} into \cref{eq:x(t)_hyp,eq:hyp_abandonment_condition}, we get \cref{eq:x(t)_hyp_2,eq:hyp_2_abandonment_condition}. 
\end{proof}

\subsection{Proof of \cref{thm:hyp_2_abandonment}}
\begin{proof}
The differentiation of the left-hand side of \cref{eq:hyp_2_abandonment_condition} is
\begin{align}
    \frac{2(kT+2)^2}{T^2} \frac{k^2 (T - t)^2 - 2k (T-t) - 2}{(k(T-t) + 2)^4}. \label{eq:hyp_2_abandonment_condition_diff}
\end{align}

(i) When $T \leq  \frac{1 + \sqrt{3}}{k}$, \cref{eq:hyp_2_abandonment_condition_diff} is non-positive in $(0, T]$. Thus, the left-hand side of \cref{eq:hyp_2_abandonment_condition} takes maximum value at $t = 0$. 

(ii) When $T > \frac{1 + \sqrt{3}}{k}$, \cref{eq:hyp_2_abandonment_condition_diff} is positive in $[0, T - \frac{1 + \sqrt{3}}{k})$, $0$ at $t = T - \frac{1 + \sqrt{3}}{k}$, and negative in $(T - \frac{1 + \sqrt{3}}{k}, T]$. Thus, the left-hand side of \cref{eq:hyp_2_abandonment_condition} takes maximum value at $t = \frac{1 + \sqrt{3}}{k}$. 
\end{proof}

\subsection{Proof of \cref{thm:exp_optimal_goal}}
\begin{proof}
As stated in \cref{subsec:abandonment_time}, $t^*$ can only take the values $0$ or $T$. In the case where $t^* = 0$, we have $x(T) = 0$. 
The condition for $t^*=T$ to hold is
\begin{align}
    \prn*{\frac{k}{(\alpha - 1) \prn*{1 - \exp \prn*{-\frac{kT}{\alpha-1}}}}}^{\alpha-1} \leq \frac{R}{\theta^{\alpha}} \iff
     \theta \leq \bra*{\frac{\alpha-1}{k} \prn*{1 - \exp \prn*{-\frac{kT}{\alpha-1}}}}^{\frac{\alpha-1}{\alpha}} R^{\frac{1}{\alpha}}, 
\end{align}
and in this case $x(T) = \theta$ holds because $t^* = T$. 
Thus, the optimal $\theta$ is $\bra*{\frac{\alpha-1}{k} \prn*{1 - \exp \prn*{-\frac{kT}{\alpha-1}}}}^{\frac{\alpha-1}{\alpha}} R^{\frac{1}{\alpha}}$, regardless of whether exploitative rewards are permitted.
\end{proof}

\subsection{Proof of \cref{thm:hyp_optimal_goal}}
\begin{proof}
(i) When $T \geq \frac{1 + \sqrt{3}}{k}$, $t^* \in [0, T - \frac{1 + \sqrt{3}}{k}]$ or $t^* = T$ hold by \cref{thm:hyp_2_abandonment}.

In the case $t^* \in [0, T - \frac{1 + \sqrt{3}}{k}]$ holds, 
\begin{align}
    &\frac{2(kT+2)^2}{T^2} \frac{(T-t^*) \prn*{k(T-t^*) + 1}}{\prn*{k(T-t^*)+2}^3} = \frac{R}{\theta^2} \\
    &\iff
    \theta = \frac{T}{kT+2} \prn*{\frac{\prn*{k(T-t^*)+2}^3 R}{2(T-t^*)\prn*{k(T-t^*) + 1}}}^{\frac{1}{2}}
\end{align}
holds from \cref{eq:hyp_2_abandonment_condition} and 
\begin{align}
    x(T) = x(t^*) &= \frac{2 t^* \theta}{(k(T - t^*) + 2) T} \\
    &= \frac{\sqrt{2}}{kT+2} \prn*{\frac{\prn*{k(T-t^*)+2}t^*{}^2 R}{(T-t^*)\prn*{k(T-t^*) + 1}}}^{\frac{1}{2}} \label{eq:x(T)_t^*}
\end{align}
from \cref{eq:x(t)_hyp_2}. 
Because we can show that \cref{eq:x(T)_t^*} is monotonically increasing for $t^* \in [0, T - \frac{1 + \sqrt{3}}{k}]$, $x(T)$ is maximized with $t^* = T - \frac{1 + \sqrt{3}}{k}$ and then $x(T) = \frac{(4 \sqrt{3} - 6) k^{\frac{1}{2}} \prn*{T - \frac{1 + \sqrt{3}}{k}} R^{\frac{1}{2}}}{k T + 2}$. 

For $t^* = T$, we have 
\begin{align}
    \frac{(kT+2)^2}{3 \sqrt{3} k T^2} \leq \frac{R}{\theta^{2}}
    \iff
    \theta \leq \frac{T(3 \sqrt{3} k)^\frac{1}{2}}{kT+2} R^{\frac{1}{2}}
\end{align}
and thus $\theta = \frac{T(3 \sqrt{3} k)^\frac{1}{2}}{kT+2} R^{\frac{1}{2}}$ is optimal and then $x(T) = \frac{T(3 \sqrt{3} k)^\frac{1}{2}}{kT+2} R^{\frac{1}{2}}$. 

Because
\begin{align}
    \frac{T \prn*{3 \sqrt{3} kR}^{\frac{1}{2}}}{kT+2} - \frac{(4 \sqrt{3} - 6) k^{\frac{1}{2}} \prn*{T - \frac{1 + \sqrt{3}}{k}} R^{\frac{1}{2}}}{k T + 2}
    = \frac{k^{\frac{1}{2}} R^{\frac{1}{2}}}{kT+2} \prn*{\prn*{6 - \sqrt{3}}T + \frac{6 - 2 \sqrt{3}}{k}} > 0
\end{align}
holds, it is optimal to make $t^* = T$ by setting $\theta = \frac{T(3 \sqrt{3} k)^\frac{1}{2}}{kT+2} R^{\frac{1}{2}}$. 
In this case, since $x(T) = \theta$ holds, this is the optimal solution regardless of whether exploitative rewards are permitted. 

(ii) When $T < \frac{1 + \sqrt{3}}{k}$, by \cref{thm:hyp_2_abandonment}, $t^*$ can  only take the value $0$ or $T$. For $t^* = 0$, we have $x(T) = 0$. 
The condition for $t^*=T$ to hold is
\begin{align}
    \frac{2(kT+1)}{kT+2} \leq \frac{R}{\theta^2}
    \iff
    \theta \leq \prn*{\frac{kT+2}{2(kT+1)}}^{\frac{1}{2}} R^{\frac{1}{2}}
\end{align}
and in this case $x(T) = \theta$ holds because $t^* = T$. 
Thus, the optimal $\theta$ is $\prn*{\frac{kT+2}{2(kT+1)}}^{\frac{1}{2}} R^{\frac{1}{2}}$, regardless of whether exploitative rewards are permitted.
\end{proof}

\subsection{Proof of \cref{thm:orsp_fixed_N_exp}}
\begin{proof}
From \cref{thm:exp_optimal_goal}, optimal $\theta_i$ is 
\begin{align}
     \theta_i = \bra*{\frac{\alpha-1}{k} \prn*{1 - \exp \prn*{-\frac{kT_i}{\alpha-1}}}}^{\frac{\alpha-1}{\alpha}} R_i^{\frac{1}{\alpha}} \label{eq:exp_optimal_theta_i}
\end{align}
given $R_i$ and $T_i$. Thus, the optimal reward scheduling problem for fixed $N$ is reduced to the problem 
\begin{align}
  \begin{aligned}
  \max_{T_i,\, R_i}\ 
  &\sum_{i=1}^N 
  \bra*{\frac{\alpha-1}{k} \prn*{1 - \exp \prn*{-\frac{kT_i}{\alpha-1}}}}^{\frac{\alpha-1}{\alpha}} R^{\frac{1}{\alpha}}_i,
  \\
  \st\quad
  &\sum_{i=1}^N T_i = T,\quad
  \sum_{i=1}^N R_i = R.
  \label{eq:problem_reward_scheduling_exp_1}
  \end{aligned}
\end{align}
Note that $T_i \in \R_{\geq 0}$ and $R_i \in \R_{\geq 0}$.
We use H\"older's inequality to upper bound the objective function of the problem \cref{eq:problem_reward_scheduling_exp_1};
\begin{align}
  \sum_{i=1}^N 
  \bra*{\frac{\alpha-1}{k} \prn*{1 - \exp \prn*{-\frac{kT_i}{\alpha-1}}}}^{\frac{\alpha-1}{\alpha}} R^{\frac{1}{\alpha}}_i 
  &\leq \bra*{\frac{\alpha-1}{k} \sum_{i=1}^N \prn*{1 - \exp \prn*{-\frac{kT_i}{\alpha-1}}}}^{\frac{\alpha-1}{\alpha}} \prn*{\sum_{i=1}^N R_i}^{\frac{1}{\alpha}} \\
  &= \bra*{\frac{\alpha-1}{k} \sum_{i=1}^N \prn*{1 - \exp \prn*{-\frac{kT_i}{\alpha-1}}}}^{\frac{\alpha-1}{\alpha}} R^{\frac{1}{\alpha}}
\end{align}
and the upper bound is achieved when
\begin{align}
    R_i = \frac{1 - \exp \prn*{-\frac{kT_i}{\alpha-1}}}{\sum_{i=1}^N \prn*{1 - \exp \prn*{-\frac{kT_i}{\alpha-1}}}}R. \label{eq:exp_optimal_R_i}
\end{align}
Therefore, the problem \cref{eq:problem_reward_scheduling_exp_1} is reduced to the following problem; 
\begin{align}
  \begin{aligned}
  \max_{T_i}\ 
  &\sum_{i=1}^N \prn*{1 - \exp \prn*{-\frac{kT_i}{\alpha-1}}},
  \\
  \st\quad
  &\sum_{i=1}^N T_i = T. 
  \label{eq:problem_reward_scheduling_exp_2}
  \end{aligned}
\end{align}
Because $1 - \exp \prn*{-\frac{kT_i}{\alpha-1}}$ is concave with respect to $T_i$. 
We can use Jensen's inequality to upper bound the objective function of \cref{eq:problem_reward_scheduling_exp_2};
\begin{align}
    \sum_{i=1}^N \prn*{1 - \exp \prn*{-\frac{kT_i}{\alpha-1}}}
    &\leq N \prn*{1 - \exp \prn*{-\frac{k \prn*{\sum_{i=1}^N \frac{T_i}{N}}}{\alpha-1}}} \\
    &= N \prn*{1 - \exp \prn*{-\frac{k T}{(\alpha-1)N}}}, 
\end{align}
and the upper bound is achieved when
\begin{align}
    T_i = \frac{T}{N}. \label{eq:exp_optimal_T_i}
\end{align}
From \cref{eq:exp_optimal_theta_i,eq:exp_optimal_R_i,eq:exp_optimal_T_i}, we obtain the optimal solution of the optimal reward scheduling problem. 
\end{proof}

\subsection{Proof of \cref{thm:orsp_fixed_N_hyp}}
\begin{proof}
From \cref{thm:hyp_optimal_goal}, optimal $\theta_i$ is 
\begin{align} \label{eq:hyp_optimal_theta_i}
     \theta_i = 
        \begin{cases}
        \frac{T_i \prn*{3 \sqrt{3} k}^{\frac{1}{2}}}{kT_i+2} R_i^{\frac{1}{2}} & T_i \geq \frac{1 + \sqrt{3}}{k}, \\
        \prn*{\frac{T_i (kT_i+2)}{2(kT_i+1)}}^{\frac{1}{2}} R_i^{\frac{1}{2}} & T_i < \frac{1 + \sqrt{3}}{k}
        \end{cases}
\end{align}
given $R_i$ and $T_i$. Thus, the optimal reward scheduling problem with fixed $N$ is reduced to 
\begin{align}
  \begin{aligned}
  \max_{T_i, R_i}\ 
  &\sum_{i=1}^N 
   g(T_i) R^{\frac{1}{2}}_i,
  \\
  \st\quad
  &\sum_{i=1}^N T_i = T,\quad
  \sum_{i=1}^N R_i = R, 
  \label{eq:problem_reward_scheduling_hyp_1}
  \end{aligned}
\end{align}
where 
\begin{align} \label{eq:definition_g}
     g(T_i) = 
        \begin{cases}
        \frac{T_i \prn*{3 \sqrt{3} k}^{\frac{1}{2}}}{kT_i+2} & T_i \geq \frac{1 + \sqrt{3}}{k}, \\
        \prn*{\frac{T_i (kT_i+2)}{2(kT_i+1)}}^{\frac{1}{2}} & T_i < \frac{1 + \sqrt{3}}{k}. 
        \end{cases}
\end{align}
Note that $T_i \in \R_{\geq 0}$ and $R_i \in \R_{\geq 0}$.
We use H\"older's inequality to upper bound the objective function of problem \cref{eq:problem_reward_scheduling_hyp_1};
\begin{align}
    \sum_{i=1}^N 
  g(T_i) R^{\frac{1}{2}}_i
  &\leq  \prn*{\sum_{i=1}^N g(T_i)^2}^{\frac{1}{2}} \prn*{\sum_{i=1}^N R_i}^{\frac{1}{2}} \\
  &= \prn*{\sum_{i=1}^N g(T_i)^2}^{\frac{1}{2}} R^{\frac{1}{2}}
\end{align}
and the upper bound is achieved when
\begin{align}
    R_i = \frac{g(T_i)^2}{\sum_{i=1}^N g(T_i)^2}R. \label{eq:hyp_optimal_R_i}
\end{align}
Therefore, \cref{eq:problem_reward_scheduling_hyp_1} is reduced to the following problem; 
\begin{align}
  \begin{aligned}
  \max_{T_i}\ 
  &\sum_{i=1}^N \tilde{g}(T_i),
  \\
  \st\quad
  &\sum_{i=1}^N T_i = T, 
  \label{eq:problem_reward_scheduling_hyp_2}
  \end{aligned}
\end{align}
where 
\begin{align}
    \tilde{g}(T_i) \coloneqq g(T_i)^2 
    = 
    \begin{cases}
        \frac{3 \sqrt{3} k T_i^2}{(kT_i+2)^2} & T_i \geq \frac{1 + \sqrt{3}}{k}, \\
        \frac{T_i (kT_i+2)}{2(kT_i+1)} & T_i < \frac{1 + \sqrt{3}}{k}. 
    \end{cases}
\end{align}
Because
\begin{align}
    \diffdiff{\tilde{g}}(T_i) 
    = 
    \begin{cases}
        \frac{24 \sqrt{3} k (1 - kT_i)}{(kT_i+2)^4} & T_i \geq \frac{1 + \sqrt{3}}{k}, \\
        -\frac{2k}{(kT_i+1)^3} & T_i < \frac{1 + \sqrt{3}}{k},
    \end{cases}
\end{align}
$\tilde{g}(T_i)$ is a concave function for $t \geq 0$. 
Thus, we can use Jensen's inequality to upper bound the objective function of the problem \cref{eq:problem_reward_scheduling_hyp_2};
\begin{align}
    \sum_{i=1}^N \tilde{g}(T_i)
    &\leq N \tilde{g}\prn*{\sum_{i=1}^N \frac{T_i}{N}} \\
    &= N \tilde{g}\prn*{\frac{T}{N}}, 
\end{align}
and the upper bound is achieved when
\begin{align}
    T_i = \frac{T}{N}. \label{eq:hyp_optimal_T_i}
\end{align}
From \cref{eq:hyp_optimal_theta_i,eq:hyp_optimal_R_i,eq:hyp_optimal_T_i}, we obtain the optimal solution of the optimal reward scheduling problem. 
\end{proof}

\subsection{Proof of \cref{thm:orsp_N_exp}}
\begin{proof}
We define $h_{\alpha}^{\mathrm{exp}}(N)$ by 
\begin{align}
h_{\alpha}^{\mathrm{exp}}(N) \coloneqq N \prn*{1 - \exp \prn*{-\frac{kT}{(\alpha-1)N}}}. 
\end{align}
Because 
\begin{align}
    \diff{h_{\alpha}^{\mathrm{exp}}}(N) = 1 - \prn*{1 + \frac{kT}{(\alpha-1)N}} \exp \prn*{-\frac{kT}{(\alpha-1)N}}, 
\end{align}
holds, we get
\begin{align}
    \lim_{N \to 0} \diff{h_{\alpha}^{\mathrm{exp}}}(N) &= 1, \\
    \lim_{N \to \infty} \diff{h_{\alpha}^{\mathrm{exp}}}(N) &= 0, \\
    \diffdiff{h_{\alpha}^{\mathrm{exp}}}(N) &= -\exp \prn*{- \frac{kT}{(\alpha-1)N}} \frac{k^2 T^2}{N^3(\alpha-1)} < 0. 
\end{align}
These formulas indicate $ \diff{h_{\alpha}^{\mathrm{exp}}}(N) > 0$ and $h_{\alpha}^{\mathrm{exp}}(N)$ is monotonically increasing for $N > 0$. 
Furthermore, we have 
\begin{align}
    \lim_{N \to \infty} h_{\alpha}^{\mathrm{exp}}(N) &= \frac{kT}{\alpha - 1} \lim_{x \to 0} \frac{1 - \exp(-x)}{x} \quad \prn*{\because x \coloneqq \frac{kT}{(\alpha-1)N}}\\
    &= \frac{kT}{\alpha - 1} \lim_{x \to 0} \frac{\exp(-x)}{1} \quad (\because \text{L'H\^opital's Rule}) \\
    &= \frac{kT}{\alpha - 1}. 
\end{align}

Because 
\begin{align}
f_{\alpha}^{\mathrm{exp}}(N) = \prn*{\frac{\alpha -1}{k} h_{\alpha}^{\mathrm{exp}}(N)}^{\frac{\alpha-1}{\alpha}} R^{\frac{1}{\alpha}}
\end{align}
holds, $f_{\alpha}^{\mathrm{exp}}(N)$ is monotonically increasing for $N > 0$ and $\lim_{N \to \infty} f_{\alpha}^{\mathrm{exp}}(N) = T^{\frac{\alpha-1}{\alpha}}R^{\frac{\1}{\alpha}}$. 
\end{proof}

\subsection{Proof of \cref{thm:orsp_N_hyp}}
\begin{proof}
Because \begin{align}
    \diff{f_2^{\mathrm{hyp}}}(N) \coloneqq
        \begin{cases}
        \frac{\prn*{3 \sqrt{3} k}^{\frac{1}{2}} T (kT - 2N)}{2 N^{\frac{1}{2}} (k T + 2N)^2} R^{\frac{1}{2}} & \frac{T}{N} \geq \frac{1 + \sqrt{3}}{k}, \\
        \frac{\sqrt{2}kT^{\frac{3}{2}}}{4 (kT + 2N)^{\frac{1}{2}} (kT+N)^{\frac{3}{2}}} R^{\frac{1}{2}} & \frac{T}{N} < \frac{1 + \sqrt{3}}{k}, 
        \end{cases} 
\end{align}
holds, 
$ \diff{f_2^{\mathrm{hyp}}}(N) > 0$ and $f_2^{\mathrm{hyp}}(N)$ is monotonically increasing for $N > 0$. Furthermore, we have 
\begin{align}
    \lim_{N \to \infty} f_2^{\mathrm{hyp}}(N)
    &= \lim_{N \to \infty} \prn*{\frac{T (kT+2N)}{2(kT+N)}}^{\frac{1}{2}} R^{\frac{1}{2}} \\
    &= T^{\frac{1}{2}} R^{\frac{1}{2}}. 
\end{align}
\end{proof}

\end{document}